\documentclass[11pt]{article}
\usepackage{graphicx}

\newsavebox{\hflrar}
\sbox{\hflrar}{\makebox[0pt][l]
{${\scriptstyle \leftharpoonup}$}{${\scriptstyle \rightharpoonup}$}}

\def \to {\rightarrow}

\begin{document}
\begin{center}
{\Large\bf Glauber Gluons in Soft Collinear Effective Theory and Factorization of Drell-Yan Processes}
\vskip 10mm
F. Liu and J.P. Ma   \\
{\small {\it Institute of Theoretical Physics, Academia Sinica, P.O. Box 2735,
Beijing 100190, China }} \\
\end{center}

\vskip 1cm
\begin{abstract}
Glauber gluons in Drell-Yan processes are soft gluons
with the transverse momenta much larger than their momentum components
along the directions of initial hadrons. Their existence has been a serious
challenge in proving the factorization of Drell-Yan processes.
The recently proposed  soft collinear effect theory of QCD can
provide a transparent way to show factorizations for a class of processes,
but it does not address the effect of glauber gluons. In this letter we first confirm the
existence of glauber gluons through an example. We then add glauber gluons
into the effective theory and study their interaction with other particles.
In the framework of the effective theory with glauber gluons we are able to show
that the effects of glauber gluons in Drell-Yan processes are canceled and the factorization
holds in the existence of glauber gluons.
Our work completes the proof or argument of factorization of Drell-Yan process
in the framework of the soft collinear effective theory.

\vskip 5mm
\noindent
\end{abstract}
\vskip 1cm
\par
QCD factorization theorems enable us to make theoretical predictions from QCD
for hard-scattering processes, where one can separate
long-distance effects from short-distance effects. The short-distance effects
depend on the details of a process and
can be studied with perturbative QCD,
while long-distance effects can be characterized by matrix elements of QCD operators,
which depend only on the structure of hadrons and infrared properties of QCD\cite{Fac}.
In general, it is difficult to prove a factorization theorem,e.g., much effort has been
spent to prove factorization for Drell-Yan process\cite{G1,G2}.
A main obstacle has been the exchange of glauber gluons in initial state interactions\cite{G0,G2}.
The glauber gluons are soft with the transverse momenta much larger than their momentum components
along the directions of initial hadrons.
\par
Recently, the soft collinear effective theory(SCET) has been proposed to study interactions among collinear-
and various soft particles of QCD\cite{SCET1,SCET2,SCETB}.
The interactions are at long-distance and produce
collinear- and I.R. divergences in perturbative calculations.
Assuming that SCET includes all those particles of QCD, then SCET will re-produce
these collinear- and I.R. singularities. It is interesting to note that
with SCET the proofs or arguments of factorization theorems can be simplified
or made in a transparent way. In \cite{SCET2} several examples including Drell-Yan processes
are given. However, the glauber gluons are not included in the proposed SCET\cite{SCET1,SCET2}.
In this letter we make an attempt to include the glauber gluons into SCET and show
that their effect in the factorization of Drell-Yan process is canceled\cite{SCET2}.
Therefore our work completes the proof or argument of factorization of Drell-Yan processes
in the framework of SCET.
\par
\begin{figure}[hbt]
\begin{center}
\includegraphics[width=10cm]{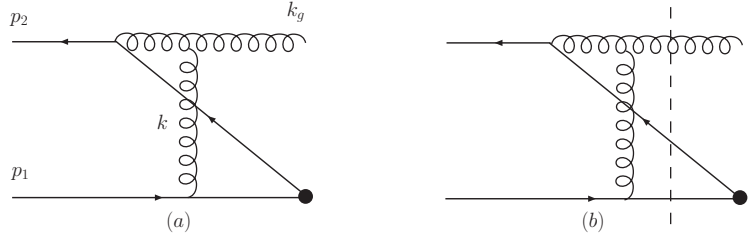}
\end{center}
\caption{A diagram which contributes to the process $q(p_1) + \bar q (p_2) \to g(k_g) + \gamma^*(q)$
at one loop level. The broken line is a cut. The black circle denotes
the insertion of the electromagnetic current operator.  }
\label{Feynman-dg2}
\end{figure}
\par
Before our study of SCET by adding glauber gluons and the factorization of Drell-Yan processes,
we show the existence of glauber gluons in scattering amplitudes. For this we consider the process
$q(p_1) + \bar q (p_2) \to g(k_g) + \gamma^*(q)$, which contributes to the differential cross section
of Drell-Yan processes. At one loop level, the amplitude of the process receives a contribution from Fig.1a.
We will use the light-cone coordinate system, in which a vector
$a^\mu$
is expressed as $a^\mu = (a^+, a^-, \vec a_\perp) = ((a^0+a^3)/\sqrt{2},
(a^0-a^3)/\sqrt{2}, a^1, a^2)$ and $a_\perp^2 =(a^1)^2+(a^2)^2$.
We introduce two light cone vectors $\bar n^\mu =(1,0,0,0)$ and $n^\mu =(0,1,0,0)$.
The momenta are given as:
\begin{equation}
p_1^\mu= (p_1^+,0,0,0), \ \ \ p_2^\mu =(0,p_2^-,0,0), \ \ \ \ k_g^\mu =(k_g^+, k_g^-, \vec k_{g\perp}).
\end{equation}
We consider the contribution in the momentum configuration where the momentum  $k_g$ scales as
$k_g^\mu  \sim Q(\lambda^2, 1,\lambda,\lambda)$
with $Q$ as a large scale and $\lambda\ll 1$. In this configuration the outgoing gluon is collinear
to the initial $\bar q$. For convenience we will take $Q=1$ in the following.
The leading contribution from Fig.1 in the limit $\lambda \to 0$ can come from different momentum regions
of the loop momentum $k$ with $k^2 \sim \lambda^2$ or with $k^2$ at higher order of $\lambda$.
These possible regions are: The collinear regions with $k$ at the order of $(1,\lambda^2,\lambda,\lambda)$
or $(\lambda^2,1,\lambda,\lambda)$, the soft region with $k$ at the order of $(\lambda,\lambda,\lambda,\lambda)$,
the ultra-soft region with $k$ at the order of $(\lambda^2,\lambda^2,\lambda^2,\lambda^2)$
and the glauber region with $k$ at the order of $(\lambda^2,\lambda^2,\lambda,\lambda)$.
Gluons with momenta in glauber regions are called as glauber gluons.
To show that  the glauber gluon gives nonzero contribution to the amplitude, it is enough
to consider the following loop integral involved in the contribution of Fig.1a:
\begin{equation}
 I = i\int \frac{d^4 k}{(2\pi)^4} \cdot \frac{1}{k^2 + i\varepsilon}\cdot
  \frac{1}{(p_1-k)^2 + i\varepsilon}  \cdot\frac{1}{(k_g-k)^2 + i\varepsilon} \cdot \frac{1}{(k-k_g+p_2)^2 + i\varepsilon}.
\end{equation}
One can perform a power counting in $\lambda$ for the integral in differen momentum regions
to determine the relative importance of contributions in deferent regions and leading contributions.
One finds that the contributions from the collinear region
with  $k \sim (\lambda^2,1,\lambda,\lambda)$, the ultra-soft region
and the glauber region  are at leading order of $\lambda$ which is ${\mathcal O}(\lambda^{-2})$.
The contributions from other regions are at higher order of $\lambda$. It is complicated
to have a clear separation of leading contributions from the three regions, because they are
all overlapped.
\par
For our purpose we can consider the absorptive part of Fig.1a instead of the total contribution. We consider
the cut diagram of Fig1.a, which is given as Fig.1b. Instead of the integral $I$ we need to consider
the absorptive- or imaginary part of $I$. The imaginary part is given by:
\begin{eqnarray}
{\rm Im} I &=&\frac{1}{2} \int \frac{d^4 k}{(2\pi)^4} \cdot \frac{1}{k^2 + i\varepsilon}\cdot
  (-2\pi i) \delta ((p_1-k)^2 )  \cdot\frac{1}{(k_g-k)^2 + i\varepsilon} (-2\pi i) \delta ((k-k_g+p_2)^2).
\end{eqnarray}
From these on-shell conditions and with $\vec k_\perp \sim (\lambda,\lambda)$
one always has $k^+ \sim \lambda^2$ and $k^- \sim \lambda^2$. This indicates
that the absorptive part of $I$ at the leading order of $\lambda$ is determined only by the glauber region.
Performing the power counting for the glauber region one obtains the leading contribution
to ${\rm Im }I$ as:
\begin{eqnarray}
{\rm Im} I_{G} &=&  \frac{1}{2} \int \frac{d^4 k}{(2\pi)^4} \cdot \frac{1}{-k^2_\perp + i\varepsilon}\cdot
  (-2\pi i) \delta (-2p_1^+ k^- -k^2_\perp )  \cdot\frac{1}{-2k_g^- k^+ -k^2_\perp + i\varepsilon}
\nonumber\\
   && \ \ \ \ \  \cdot (-2\pi i) \delta (-2 k^+ (k_g^--p_2^-)
   + 2 \vec k_{g\perp} \cdot \vec k_\perp -k_\perp^2 +(k_g-p_2)^2)
\nonumber\\
   &=& - \frac{1}{8 p_1^+ p_2^-} \int \frac{d^2 k_\perp}{(2\pi)^2} \frac{1}{k^2_\perp (\vec k_\perp -\vec k_{g\perp})^2}
 =  \frac{1}{8 \pi s k^2_{g\perp} } \left [
   \left ( \frac{2}{\epsilon} -\gamma + \ln 4\pi \right ) - \ln \frac{k^2_{g\perp} }{\mu^2} \right] +{\mathcal O}(\lambda^0),
\end{eqnarray}
with $s=2p_1^+ p_2^-$. In the above we have used dimensional regularization with $\epsilon =4-d$
to regularize divergences.
From the result the integral is divergent. The divergent contribution comes partly
form the region of $\vec k_{\perp} \sim 0$ and partly from the region of $\vec k_\perp \sim \vec k_{g\perp}$.
The singular contribution from the region of $\vec k_\perp \sim \vec k_{g\perp}$
is because that there are pinched poles in the complex $k^+$-plan for the integral $I$.
The singularities generated by these pinched poles have brought up a difficulty to prove the factorization
in Drell-Yan processes.
\par
From the above discussion we have
${\rm Im }I = {\rm Im} I_G( 1+ {\mathcal O}(\lambda))$.  To verify this one can work out
the exact result of the integral $I$. The exact result can be found in \cite{INT}.
Through an analytical continuation and an expansion of the exact result we have from \cite{INT}:
\begin{equation}
{\rm Im } I =\frac{1}{8 \pi s k^2_{g\perp} } \left [
   \left ( \frac{2}{\epsilon} -\gamma + \ln 4\pi \right ) - \ln \frac{k^2_{g\perp} }{\mu^2} \right]
   + {\mathcal O} ( k_{g\perp}^0 ).
\end{equation}
Comparing the result in Eq.(4,5) one verifies ${\rm Im }I = {\rm Im} I_G( 1+ {\mathcal O}(\lambda))$.
One can conclude that the exchange of glauber gluons does give nonzero contributions
to scattering amplitudes at leading order of $\lambda$ and hence possibly to differential cross sections.
The possible contributions due to glauber gluons are at the same order of contributions
due to collinear- and ultra-soft gluons. Therefore, one should include glauber gluons
into SCET as an effective theory of QCD to describe interactions between collinear partons.
Although the contributions of glauber gluons in Drell-Yan processes at twist-2 level are canceled,
as shown in \cite{G1,G2} and in the current work, but the existence of glauber gluons
has observable effects at twist-3 level. Recent studies of single transverse spin asymmetries
in Drell-Yan- and DIS processes have been shown that these asymmetries are generated through
exchanges of glauber gluons\cite{SSA}. Hence it is also important to have SCET containing glauber
gluons for a possible study of factorizations of single transverse spin asymmetries.
\par
To derive SCET containing glauber gluons, we first consider SCET of one collinear system, later
we will add another collinear system for our purpose. We will use the space-time representation
of SCET as used in \cite{SCETB}.
We choose the collinear system moving along the $\bar n$-direction with the momentum $p$ at
the order
\begin{equation}
 p^\mu \sim Q(1,\lambda^2, \lambda, \lambda),
\end{equation}
where $Q$ is a larger scale while $\lambda$ is small with $\lambda\ll 1$.
Again we will take $Q=1$ in the following. Interacting particles in  such a collinear system
can be classified with their momenta at different orders of $\lambda$. They are collinear particles
with momenta at the order $p^\mu \sim (1,\lambda^2, \lambda, \lambda)$ , ultra-soft particles
with momenta at the order $p^\mu\sim(\lambda^2,\lambda^2, \lambda^2, \lambda^2)$, and glauber gluons
with momenta at the order $p^\mu\sim(\lambda^2,\lambda^2, \lambda, \lambda)$. The later have not been incorporated
in SCET.
Soft particles with momenta at the order of $p^\mu \sim (\lambda,\lambda, \lambda, \lambda)$
will not interact with the above particles because of momentum conservation. However,
their effect will appear in matching operators of full QCD into those of SCET. It has been shown
that their effect is canceled in Drell-Yan process. We will not consider soft particles here.
\par
For the collinear system
we decompose the gluon field and quark field of full QCD as:
\begin{eqnarray}
   A^\mu = A_{\bar n}^\mu + A_g^\mu + A_{us}^\mu, \ \ \ \ \
   \psi = \xi_{\bar n} + \eta + q.
\end{eqnarray}
In the above $A_{\bar n}$ and $\xi_{\bar n}$  are the collinear fields for collinear gluons
 and collinear quarks, respectively.  $\eta$ is a small component for collinear quarks
which will be integrated out and expressed in terms of other fields.
$A_{us}$ and $q$ are the ultra-soft fields  for ultra-soft particles.
$A_g^\mu$ is the field for glauber gluons. The space-time dependence of these fields are characterized
by the small scale $\lambda$ in the following patten. For a collinear field $\phi_{\bar n}$ which can be
the collinear gluon field or the collinear quark field we have
\begin{equation}
 \partial^+ \phi_{\bar n}(x) \sim {\mathcal O}(1) \phi_{\bar n}(x), \ \
 \partial^-\phi_{\bar n}(x) \sim {\mathcal O}(\lambda^2) \phi_{\bar n}(x), \ \
 \partial_\perp^\mu \phi_{\bar n}(x) \sim {\mathcal O}(\lambda) \phi_{\bar n}(x).
\end{equation}
For a generic ultra-soft field $\phi_{us}$  and the glauber gluon field we have
\begin{eqnarray}
   \partial^\mu \phi_{us}(x) & \sim &  {\mathcal O}(\lambda^2) \phi_{us}(x),
\nonumber\\
   \partial^+ A_g^\mu (x) & \sim & {\mathcal O}(\lambda^2)A_g^\mu (x), \ \
 \partial^- A_g^\mu (x) \sim {\mathcal O}(\lambda^2) A_g^\mu (x), \ \
 \partial_\perp^\nu A_g^\mu (x) \sim {\mathcal O}(\lambda) A_g^\mu (x).
\end{eqnarray}
By inspecting propagators of particles in different momentum modes we can derive
the following power-scaling:
\begin{eqnarray}
A^+_{\bar n} &\sim & {\mathcal O}(1), \ \ \ A^-_{\bar n} \sim  {\mathcal O}(\lambda^2), \ \ \
A^\mu_{\bar n \perp } \sim  {\mathcal O}(\lambda), \ \ \ \xi \sim {\mathcal O}(\lambda),
\nonumber\\
A^\mu_{us} & \sim &   {\mathcal O}(\lambda^2), \ \ \  A^\mu_{g}  \sim    {\mathcal O}(\lambda^2),
\ \ \
q \sim  {\mathcal O}(\lambda^3 ).
\end{eqnarray}
\par
Using the power-scaling of fields and the scaling of the space-time dependence of the fields one can expand
the Lagrangian of full QCD  in $\lambda$ with the above fields. The expansion is straightforward.
To express the leading order results we introduce the modified derivative $\partial^\mu_{\bar n}$ or projection.
The derivative $\partial^+_{\bar n}$ only acts on collinear fields, it gives zero if the derivative acts
on ultra-soft- or glauber gluon fields, i.e.,
\begin{equation}
\partial_{\bar n}^+ \phi_{\bar n} (x) =\partial^+ \phi_{\bar n} (x), \ \ \
\partial_{\bar n}^+ \phi_{us} (x) = 0, \ \ \ \partial_{\bar n}^+ A_g^\mu =0.
\end{equation}
The derivative $\partial^-_{\bar n}$ acts as an usual derivative on all fields while $\partial^\mu_{\bar n \perp}$
acts only on collinear- or glauber gluon fields, but not on ultra-soft fields. The leading terms of
our effective theory can be written as:
\begin{equation}
{\mathcal L}^{(\bar n)} (\xi, q, A_{\bar n}^\mu, A_g^\mu, A_{us}^\mu ) = {\mathcal L}_g (A_g^\mu)
 + {\mathcal L} _{us}(q, A_{us}^\mu ) + {\mathcal L}_{\bar n} (\xi_{\bar n}, A_{\bar n}^\mu, \bar n\cdot A_g, \bar n \cdot A_{us}) +\cdots,
\end{equation}
where $\cdots$ stand for terms whose  contributions to the action $S=\int d^4 x {\mathcal L}^{(\bar n)}$ are suppressed by $\lambda$.
In the expansion of the action $S$ one should assign the space-time integration $d^4 x$ with
a correct scaling in $\lambda$ by noticing that the integration should not change
the momentum orders of fields in different momentum modes.
\par
At leading order glauber gluons can not interact with ultra-soft particles and they also can not interact
with themselves. Hence ${\mathcal L}_g (A_g)$ is only the kinetic term of the glauber gluons:
\begin{equation}
 {\mathcal L}_g = -\frac{1}{2} {\rm Tr} \big [ \partial_\perp ^\mu A_g^\nu - \partial_\perp^\nu A_g^\mu \big]
                              \big [ \partial_{\perp \mu} A_{g\nu} - \partial_{\perp\nu} A_{g \mu} \big ].
\end{equation}
It should be noted that only the transverse derivative appears. This fact results in that
the propagator of glauber gluons with the momentum $k$ will be proportional to $1 / k^2_\perp$,
as already seen in Eq.(4).
${\mathcal L}_{us}(q, A_{us}^\mu )$ is the part for ultra-soft particles:
\begin{equation}
 {\mathcal L}_{us}(q, A_{us}^\mu ) = \bar q i \gamma \cdot D_{us}  q  -\frac{1}{2g^2} {\rm Tr } \left \{
                \left [ D_{us}^\mu, D_{us}^\nu \right ] \right \}^2,
\end{equation}
where the covariant derivative is defined only with the ultra-soft gluon field, i.e., $D_{us}^\mu =\partial^\mu
 + ig A^\mu_{us}$.  ${\mathcal L}_{\bar n} (\xi_{\bar n}, A_{\bar n}, A_g, A_{us})$ contains collinear particles
 and their interactions with ultra-soft- and glauber gluons. To express it we introduce
 the covariant derivative with the modified derivative:
 \begin{equation}
    D_{\bar n}^\mu =\partial_{\bar n}^\mu + ig A_{\bar n} ^\mu + ig n^\mu \bar n \cdot (A_g + A_{us}) .
\end{equation}
With the covariant derivative the part ${\mathcal L}_{\bar n}$ can be expressed as
\begin{eqnarray}
{\mathcal L} _{\bar n} (\xi_{\bar n}, A_{\bar n}^\mu, \bar n\cdot A_g, \bar n \cdot A_{us}) & = &
 \bar \xi_{\bar n} i n\cdot \gamma \bar n \cdot D_{\bar n} \xi_{\bar n} - \bar\xi_{\bar n} i\gamma\cdot D_{c\perp}
  \frac{1}{ 2iD_c^+} n\cdot \gamma  i\gamma\cdot D_{c\perp} \xi_{\bar n}
\nonumber\\
   &&   -\frac{1}{2 g^2} {\rm Tr } \left \{
                \left [ D_{\bar n }^\mu, D_{\bar n}^\nu \right ] \right \}^2,
\end{eqnarray}
where the collinear covariant derivative $D_c$ is defined as $D_c^\mu =\partial^\mu + ig A^\mu_{\bar n}$.
The glauber gluon field appears in the covariant derivative in the same way as the ultra-soft gluon field.
However, the interaction of collinear gluons with ultra-soft gluons are different than that
with glauber gluons because the operator $\partial_{\bar n \perp}$ does not act on the ultra-soft field, but
 on the glauber field, i.e.,  $\partial_{\bar n \perp}^\mu A_{g} \neq 0$.
\par
The effective theory at leading order $\lambda$ is invariant under various gauge transformations,
which are collinear-, ultra-soft- and glauber gauge transformation. The space-time dependence of these
transformations is characterized in the same patten as that of the corresponding fields in Eq. (8,9).
These transformations are:
\begin{eqnarray}
{\rm Collinear:}\  A_{us}^\mu &\to & A_{us}^\mu, \ \ \ \  A_g^\mu \to A_g^{\mu}, \ \ \ \
           \xi_{\bar n} \to U_c \xi_{\bar n}, \ \ \ \ \  q \to q,
\nonumber\\
                   A^\mu_{\bar n} &\to& U_c A_{\bar n}^\mu U_c^\dagger -\frac{i}{g} U_c \left [ \partial^\mu
                       +ig(A^\mu_{us} + A^\mu_{g} ), U_c^\dagger \right ],
\nonumber\\
{\rm Ultra-soft:} \  A_{\bar n}^\mu &\to & U_{us}  A_{\bar n}^\mu U^\dagger_{us}, \ \ \ \
                    A_{g}^\mu  \to  U_{us}  A_{g}^\mu U^\dagger_{us},  \ \ \ \ \xi_{\bar n} \to U_{us} \xi_{\bar n},
                     \ \ \ \  q\to U_{us} q,
\nonumber\\
                   A^\mu_{us} & \to & U_{us} A_{us}^\mu U_{us}^\dagger -\frac{i}{g} U_{us} \left [ \partial^\mu
                   , U_{us}^\dagger \right ],
\nonumber\\
{\rm Glauber:} \     A_{g}^\mu  & \to  &   A_{g}^\mu, \ \ \ \ A^\mu_{us} \to A^\mu_{us}, \ \ \ \
                   \xi_{\bar n} \to U_{g} \xi_{\bar n},
                     \ \ \ \  q\to  q,
\nonumber\\
                   A^\mu_{\bar n} & \to & U_g A_{\bar n}^\mu U_g^\dagger -\frac{i}{g} U_{g} \left [ \partial^\mu
                   +ig(A^\mu_{us} + A^\mu_{g} ), U_g^\dagger \right ].
\end{eqnarray}
Under these transformations the effective Lagrangian is invariant at the leading order of $\lambda$, or
the change of the effective Lagrangian is suppressed by $\lambda$. One can re-express these transformations
with the modified derivative and neglect higher order effects introduced by the fields in the
transformation so that
the effective Lagrangian is exactly invariant. In the above glauber gauge transformation
the glauber field is not changed. However, there is another type of transformations
related to the glauber transformation defined by $U_{g\perp}(\tilde x) = U_g(x^+ \bar n + x^- n)$, under which
the glauber field is changed:
\begin{eqnarray}
    A_{g}^\mu  & \to  &   U_{g\perp}  A_{g}^\mu U_{g\perp}^\dagger -\frac{i}{g} U_{g\perp}
            \left [\partial^\mu , U_{g\perp}^\dagger \right ],
\ \ \ \ A^\mu_{us} \to A^\mu_{us},
                     \ \ \ \  q\to  q,
\nonumber\\
                   A^\mu_{\bar n} & \to & U_{g\perp}  A_{\bar n}^\mu U_{g\perp}^\dagger + U_{g\perp} \left [
                   A^\mu_{us}, U_{g\perp}^\dagger \right ], \ \ \ \
                   \xi_{\bar n} \to U_{g\perp} \xi_{\bar n}.
\end{eqnarray}
Under the above transformation the Lagrangian is invariant. One needs to fix gauges for different gauge fields
and introduces corresponding ghost fields. We note that the gauge fixing for ultra-soft- and glauber gluons
can be fixed separately. We will take Feynman gauge for ultra-soft- and glauber gluons. For collinear
gluons we take the covariant gauge for collinear gluons as in \cite{SCET1}, where the covariant derivative
should include the glauber gluon field as in Eq.(15).
\par
To study the effects of glauber- and ultra-soft gluons in Drell-Yan process one needs to include
another collinear fields in the $n$-direction. This can be done by adding the collinear gluon
field $A_n^\mu$ and the collinear quark field $\xi_n$ in the decomposition in Eq.(7) correspondingly
and by doing the expansion in $\lambda$. At leading order $\lambda$ the collinear fields in different
directions can no interact with each other. Momentum conservation insures that
the collinear fields in different directions can not be mixed with each other.
Then, the soft collinear effective
theory for two collinear systems moving in the $n$-direction and in $\bar n$ direction respectively
can be easily written down:
\begin{equation}
{\mathcal L}^{(\bar n n)} = {\mathcal L}_g (A_g^\mu) + {\mathcal L}_{us}(q, A_{us}^\mu )
 + {\mathcal L} _{\bar n} (\xi_{\bar n}, A_{\bar n}^\mu, \bar n\cdot A_g, \bar n \cdot A_{us}) +
 {\mathcal L}_{ n} (\xi_{ n}, A_n^\mu,  n\cdot A_g, n \cdot A_{us})+ \cdots.
\end{equation}
From the effective theory, the interaction between the collinear fields
in different directions is only through exchanging glauber- and ultra-soft gluons.
\par
Without the presence of glauber gluons the decoupling of ultra-soft gluons
can be simply made through redefinitions of collinear fields, as shown in \cite{SCET1,SCET2}.
In the presence of glauber gluons in SCET, one can also redefine fields so that
the redefined fields do not interacting with glauber gluons and ultra-soft gluons.
We introduce the following $SU(3)$-matrices:
\begin{eqnarray}
    Z_{\bar n} (x) &=&  P \exp \left \{ -ig \int_{-\infty}^0 d\beta \bar n
    \cdot  A_g (\beta \bar n + x) \right \},\ \
    Z_{ n} ( x) =  P \exp \left \{ -ig \int_{-\infty}^0 d\beta  n
    \cdot A_g (\beta n + x)   \right \},
\nonumber\\
   Y_{\bar n} (x) &=&  P \exp \left \{ -ig \int_{-\infty}^0 d\beta \bar n
    \cdot  A_{us} (\beta \bar n +x) \right \},\ \
    Y_{ n} ( x) =  P \exp \left \{ -ig \int_{-\infty}^0 d\beta  n
    \cdot A_{us} (\beta n + x)   \right \}.
\nonumber\\
\end{eqnarray}
With these matrices we redefine all collinear fields as:
\begin{eqnarray}
A_{\bar n}^\mu &=& Z_{\bar n} \left [ Y_{\bar n} \tilde A^\mu_{\bar n} Y^\dagger_{\bar n} \right ] Z^\dagger_{\bar n}
   -\frac{i}{g} Z_{\bar n} \left [ \partial_\perp^\mu + ig_s n^\mu A_{us}^-, Z^\dagger_{\bar n}\right ], \ \ \ \
   \xi_{\bar n} = Z_{\bar n} Y_{\bar n} \chi_{\bar n},
\nonumber\\
A_{n}^\mu &=& Z_{n} \left [ Y_{n} \tilde A^\mu_{n} Y^\dagger_{n} \right ] Z^\dagger_{ n}
   -\frac{i}{g} Z_n \left [ \partial_\perp^\mu + ig_s \bar n^\mu A_{us}^+, Z^\dagger_{n}\right ], \ \ \ \
   \xi_{ n} = Z_{ n} Y_{ n} \chi_{n}.
\end{eqnarray}
With the re-defined fields the Lagrangian containing the interactions with glauber gluons
at the leading order of $\lambda$ becomes
\begin{equation}
{\mathcal L} _{\bar n} (\xi_{\bar n}, A_{\bar n}^\mu, \bar n\cdot A_g , \bar n\cdot A_{us} )
  =  {\mathcal L}_{\bar n} (\chi_{\bar n}, \tilde A_{\bar n}^\mu, 0, 0), \ \ \ \
{\mathcal L} _{n} (\xi_{n}, A_{n}^\mu,  n\cdot A_g, n\cdot A_{us})
  =  {\mathcal L}_{n} (\chi_{ n}, \tilde A_{ n}^\mu, 0, 0).
\end{equation}
This result tells that the redefined collinear fields will not interact with glauber gluons
and ultra-soft gluons. It should be noted that there is no difference between $S$-matrix elements calculated
with the original Lagrangian and those calculated with the Lagrangian of redefined fields.
For our redefinition of fields this can also be realized with path-integral of SCET by noting
the fact that the Jacobian's of changing field variables are unit.
\par
Now we are in position to discuss the factorization for Drell-Yan process. The process is given as
\begin{equation}
     h_{\bar n} + h_{n} \to \ell^+ + \ell^- + X,
\end{equation}
where one hadron $h_n$ moves in the $n$-direction while another $h_{\bar n}$
moves in the $\bar n$-direction. At leading order
of QED the differential cross-section is determined by the hadronic tensor:
\begin{equation}
 W^{\mu\nu}(q) = \int d^4 x e^{-i q\cdot x} \langle h_{\bar n} h_n \vert
     J^\mu (x) J^\nu (0) \vert  h_n  h_{\bar n} \rangle, \ \ \  J^\mu =\bar \psi \gamma^\mu \psi .
\end{equation}
To show the factorization one needs first to match the operator $ J^\mu (x) J^\nu (0)$ to the operators
of the soft collinear effective theory at the leading order in $\lambda$. In the matching, the interactions
between collinear particles which are not presented in SCET are assembled by different Wilson lines:
\begin{equation}
    W_{\bar n } (x) = P\exp \left \{ -ig \int_{-\infty}^0 d\alpha n\cdot A_{\bar n} ( \alpha n +x) \right \},
  \ \ \ \ W_{n } (x) = P\exp \left \{ -ig \int_{-\infty}^0 d\alpha \bar n\cdot A_{n} ( \alpha  \bar n +x) \right \}.
\end{equation}
At leading order of $\lambda$ the operators consists of collinear fields only and their relative space-time dependence
can only be of $x^-n$ for $\phi_{\bar n}$ fields and of $x^+ n$ of $\phi_n$. Taking the operators
consisting only of quark fields in SCET as example, the matching is:
\begin{eqnarray}
W^{\mu\nu}(q) &=& \int dk_1^+ d k_2^- C^{\mu\nu}_{ij}(q, k_1^+ \bar n, k_2^- n )
                         \int d x^- dy^+ \exp \left (-i x^- k_1^+ -i y^+ k_2^- \right )
\nonumber\\
          &&  \cdot \langle h_{\bar n} h_n \vert
     \left [ ( \bar   \xi_{\bar n} W_{\bar n} )(x^- n) \Gamma_{\bar n}^{(i)} (W_{\bar n}^\dagger \xi_{\bar n} ) (0) \right ]
     \left [  ( \bar   \xi_{ n} W_ n ) (y^+ \bar n )\Gamma_{n}^{(j)} (W_n^\dagger \xi_{ n} ) (0) \right ]
        \vert  h_n  h_{\bar n} \rangle
     +\cdots ,
\end{eqnarray}
where $\cdots$ stand for other possible operators at leading order and terms at higher
orders in $\lambda$ which will be neglected. The leading order is at $\lambda^4$.
For simplicity we will discuss the factorization for the above term in detail. Those terms
containing other possible operators can be handled in the same way.
In the above we have already
used the properties
\begin{equation}
  \bar n \cdot \gamma \xi_{\bar n} = \gamma^- \xi_{\bar n}=0, \ \ \ \
   n \cdot \gamma \xi_{n} = \gamma^ + \xi_{n}=0
\end{equation}
to decouple the Dirac indices. Hence the $\Gamma_{\bar n} $-matrices are given by
\begin{equation}
  \left ( \Gamma_{\bar n}^{(1)}, \Gamma_{\bar n}^{(2)},\Gamma_{\bar n}^{(3)} \right )
     = \left ( n\cdot \gamma,  n\cdot \gamma \gamma_5, n\cdot\gamma \gamma^\nu_\perp, \right ).
\end{equation}
$\Gamma_n^{(i)}$ is defined by replacing $n$ with $\bar n$ in $\Gamma_{\bar n}^{(i)}$.
\par
The functions $C$ can be calculated with perturbative theory.
If one uses perturbative theory to calculate the hadronic tensor by replacing the hadrons with parton states,
one will have I.R.- and collinear singularities. These singularities are reproduced by the matrix elements
of SCET operators through construction of SCET, e.g., those in the right hand side of the above equation. Hence
the perturbative functions are free from I.R.- and
collinear divergence. At this stage the long-distance- and short-distance effects
are factorized, but the factorization of Drell-Yan process is not
complete.
The factorization is completely achieved if one can show that the matrix element
of the two-hadron state in Eq.(26)
can be factorized into a product of  two matrix elements of one-hadron state.
In Eq.(26) the operator $\left [ ( \bar   \xi_{\bar n} W_{\bar n} )(x^- n) \Gamma_{\bar n}^{(i)} (W_{\bar n}^\dagger \xi_{\bar n} ) (0) \right ]$
can emit ultra-soft- and glauber gluons to interact with the state $h_n$ and the emitted
gluons can also be absorbed by the operator
$\left [  ( \bar   \xi_{ n} W_ n ) (y^+ \bar n )\Gamma_{n}^{(j)} (W_n^\dagger \xi_{ n} ) (0) \right ]$.
This prevents us at first look from factorizing the matrix element of two-hadron state in Eq.(26)
as a product of two matrix elements of one-hadron state.
For the factorization one needs to show that the total effects of gluon-emissions are canceled.
\par
Before discussing the factorization, we discuss a general feature of the $\lambda$-expansion
in the space-time representation. We consider a product of collinear fields $\phi_{\bar n}$, denoted
as $\Phi_{\bar n}$, multiplied with any ultra-soft field $\phi_{us}$. The integral of the product
has the following property:
\begin{equation}
\int d\beta  \Phi_{\bar n} (\beta n +x) \phi_{us} (\beta n +x) = \left [ \int d\beta
\Phi_{\bar n} (\beta n +x) \right ]  \phi_{us} ( x) \left \{ 1 + {\mathcal O}(\lambda^2) \right \}.
\end{equation}
This equation can be derived by expansion the ultra-soft field as $ \phi_{us} (\beta n +x) =  \phi_{us} ( x)
+ \beta \partial^+\phi_{us}(x) +\cdots$. The derivative $ \partial^+\phi_{us}(x)$ is at the order of $\lambda^2$,
while the integration variable $\beta$ is related to the collinear direction and is taken as at the order
of $\lambda^0$. Therefore we have the above equation. Now we  use the redefined
collinear fields in Eq.(21) to express the matrix elements in Eq.(26).
Under the re-definition the Wilson $W_{n}$ and $W_{\bar n}$ become by using the property
\begin{eqnarray}
W_n (x) &=& V_{n}(x)  \tilde W_n (x) V_{n}^\dagger (x)
\left \{ 1 +{\mathcal O}(\lambda^2)\right \},\ \ \ \  V_n (x) =Z_{n}(x) Y_n ( x ),
\nonumber\\
W_{\bar n} (x) &=&  V_{\bar n}(x)  \tilde W_{\bar n} (x)
V_n^\dagger (x)
\left \{ 1 +{\mathcal O}(\lambda^2)\right \}, \ \ \ \ V_{\bar n} (x) =Z_{\bar n}(x) Y_{\bar n} ( x )
\end{eqnarray}
where $\tilde W_{n}$ and $\tilde W_{\bar n}$ are defined by replacing $A_n$ and $A_{\bar n}$ in Eq.(25)
with $\tilde A_n$ and $\tilde A_{\bar n}$, respectively.
The hadronic matrix element with the redefined fields become:
\begin{eqnarray}
 && \langle h_{\bar n} h_n \vert
  \left [ ( \bar   \xi_{\bar n} W_{\bar n} )(x^- n) \Gamma_{\bar n}^{(i)} (W_{\bar n}^\dagger \xi_{\bar n} ) (0) \right ]
     \left [  ( \bar   \xi_{ n} W_ n ) (y^+ \bar n )\Gamma_{n}^{(j)} (W_n^\dagger \xi_{ n} ) (0) \right ]
        \vert  h_n  h_{\bar n} \rangle
\nonumber\\
     && =  \langle h_{\bar n}  h_n \vert
      ( \bar \chi_{\bar n} \tilde W_{\bar n} )_i (x^- n)
       \Gamma_{\bar n}^{(i)} (\tilde W_{\bar n}^\dagger \chi_{\bar n} )_j  (0)
 \left [ V_{\bar n}^\dagger (x^-n) V_{\bar n} (0) \right ]_{ij}
                 \left [ V_{n}^\dagger (y^+ \bar n ) V_{n} (0) \right ]_{kl}
\nonumber\\
       && \ \ \ \ \ \ \ \ \ \ \ \ \  ( \bar   \chi _{ n} \tilde W_ n )_k (y^+ \bar n )\Gamma_{n}^{(j)}
      (\tilde W_n^\dagger \chi_{ n} )_l(0)
        \vert  h_n h_{\bar n} \rangle \left [ 1 + {\mathcal O}(\lambda^2)\right ] ,
\end{eqnarray}
where $ijlk$ are color indices. With the definition of $V_{\bar n}(x^-n)$ and
$V_{n}(y^+ \bar n)$ one knows that the $x^-$-dependence in $V_{\bar n}(x^-n)$
and the $y^+$-dependence in $V_{n}(y^+ \bar n)$ are characterized by $\lambda^2$.
Since the space-time separation $x^-$ and $y^+$ will be integrated over in the
hadronic tensor, hence we can have the following approximation under the integrations:
\begin{eqnarray}
 \left [ V_{\bar n}^\dagger (x^-n) V_{\bar n} (0) \right ]_{ij}
                 \left [ V_{n}^\dagger (y^+ \bar n ) V_{n} (0) \right ]_{kl}
              = \delta_{ij}\delta_{lk} + {\mathcal O}(\lambda^2).
\end{eqnarray}
Using the fact that redefined collinear fields do not interact with glauber- and ultra-soft gluons and also
that the collinear fields in different directions do not interact, we can re-write Eq.(31) under the integrations
in Eq.(26) with Eq.(32) as:
\begin{eqnarray}
 && \langle h_{\bar n} h_n \vert
  \left [ ( \bar   \xi_{\bar n} W_{\bar n} )(x^- n) \Gamma_{\bar n}^{(i)} (W_{\bar n}^\dagger \xi_{\bar n} ) (0) \right ]
     \left [  ( \bar   \xi_{ n} W_ n ) (y^+ \bar n )\Gamma_{n}^{(j)} (W_n^\dagger \xi_{ n} ) (0) \right ]
        \vert  h_n  h_{\bar n} \rangle
\nonumber\\
     && =  \left [ \langle h_{\bar n} \vert
      ( \bar \chi_{\bar n} \tilde W_{\bar n} )(x^- n)
       \Gamma_{\bar n}^{(i)} (\tilde W_{\bar n}^\dagger \chi_{\bar n} )  (0) \vert h_{\bar n} \rangle \right ]
  \left [ \langle h_n ( \bar   \chi _{ n} \tilde W_ n ) (y^+ \bar n )\Gamma_{n}^{(j)}
      (\tilde W_n^\dagger \chi_{ n} )(0)
        \vert  h_n \rangle \right ]
\nonumber\\
   && \ \ \cdot \left \{ 1 + {\mathcal O}(\lambda^2)\right \}.
\end{eqnarray}
Substituting this result and expressing the redefined fields with the original fields, we complete
the factorization for Eq. (26):
\begin{eqnarray}
W^{\mu\nu}(q) &=& \int dk_1^+ d k_2^- C^{\mu\nu}_{ij}(q, k_1^+ \bar n, k_2^- n )
 \cdot \left [ \int d x^- e^{-i x^- k_1^+ } \langle h_{\bar n} \vert
      ( \bar   \xi_{\bar n} W_{\bar n} )(x^- n) \Gamma_{\bar n}^{(i)} (W_{\bar n}^\dagger \xi_{\bar n} ) (0)
      \vert h_{\bar n} \rangle \right ]
\nonumber\\
       && \cdot \left [ \int dy^+ e^{  -i y^+ k_2^- }
     \langle h_n \vert    ( \bar   \xi_{ n} W_ n ) (y^+ \bar n )\Gamma_{n}^{(j)} (W_n^\dagger \xi_{ n} ) (0)
        \vert  h_n  \rangle \right ]
  \left [ 1 +{\mathcal O}(\lambda^2)  \right ]  + \cdots.
\end{eqnarray}
In the matching in Eq.(26) there are contributions from color-octet operators at leading order
like
\begin{equation}
\left [ ( \bar   \xi_{\bar n} W_{\bar n}  )(x^- n)  \Gamma_{\bar n}^{(i)}T^a
      ( W_{\bar n}^\dagger \xi_{\bar n} ) (0) \right ]
      \left [  ( \bar   \xi_{ n} W_ n  )(y^+ \bar n ) \Gamma_{n}^{(j)}T^a  (W_n^\dagger \xi_{ n} )  (0) \right ],
\end{equation}
represented by $\cdots$. With the above procedure and the fact that the hadrons are color-less, one can show
that these color-octet operators do not contribute at leading order of $\lambda$.
\par
In the above we have shown
that the effects of glauber gluons are completely canceled at the leading order of $\lambda$ in the same
way of the cancelation of ultra-soft gluons. The above result is invariant
at leading order of $\lambda$ under various gauge transformation. We note that the Wilson line
$W_{n}$ and $W_{\bar n}$ under the transformations given in Eq. (17) are transformed as:
\begin{eqnarray}
{\rm Collinear:}\  W_n &\to & U_n W_n \left ( 1+{\mathcal O}(\lambda^2)\right ), \ \ \
                   W_{\bar n} \to U_{\bar n} W_{\bar n} \left ( 1+{\mathcal O}(\lambda^2)\right ).
\nonumber\\
{\rm Ultra-soft:} \   W_n &\to & U_{us} W_n U_{us}^\dagger \left ( 1+{\mathcal O}(\lambda^2)\right ), \ \ \
                   W_{\bar n} \to U_{us} W_{\bar n} U_{us}^\dagger \left ( 1+{\mathcal O}(\lambda^2)\right ).
\nonumber\\
{\rm Glauber:} \      W_n &\to & U_{g} W_n U_{g}^\dagger \left ( 1+{\mathcal O}(\lambda^2)\right ), \ \ \
                   W_{\bar n} \to U_{g} W_{\bar n} U_{g}^\dagger \left ( 1+{\mathcal O}(\lambda^2) \right ).
\end{eqnarray}
At the leading order of $\lambda$ there are other three operators appearing in the matching in Eq.(26).
They are:
\begin{equation}
\bar \xi_n \xi_n G^{+\mu}_{\bar n} G^{+\nu}_{\bar n}, \ \ \ \
\bar \xi_{\bar n} \xi_{\bar n} G^{-\mu}_{ n} G^{-\nu}_{n}, \ \ \ \
G^{-\mu}_{ n} G^{-\nu}_{n} G^{+\mu'}_{\bar n} G^{+\nu'}_{\bar n},
\end{equation}
where we have suppressed the color structure,  spin structures, space-time dependence
and the Wilson line of collinear gluon in different directions. The collinear field
strength tensors are defined with corresponding collinear gluon fields, the indices
$\mu,\nu, \mu'$ and $\nu'$ are transverse. The factorization or the cancelation of
effects through exchanges of glauber- and ultra-soft gluons can be shown in a similar way
as in the above. With the factorization one can define for each hadron
various parton distributions and write the final result for the hadronic tensor
in a compact form. The detailed results can be found somewhere in the literature.
\par
To summarize: In this letter we have studied the effects of glauber gluons in the
framework of SCET, which had not been considered in the originally proposed SCET.
We have confirmed the existence of glauber gluons through an example.
By adding the glauber gluons into SCET we have found that the glauber gluons only
interact with collinear particles. Furthermore, we have shown in the framework of SCET
with glauber gluons that the effects of glauber gluons are canceled in Drell-Yan process,
i.e., the factorization in the present of glauber gluons still holds.
Therefore our work completes the proof or argument of factorization of Drell-Yan process
in the framework of SCET.
Our result
is in the agreement with the traditional proof of Drell-Yan process where
the existence of glauber gluon has been a challenging difficulty. However
the proof here is given in a more transparent way.

\vskip 5mm
\par\noindent
{\bf\large Acknowledgments}
\par
We thank Prof. M. Yu and Dr. F. Feng for many interesting discussions.
This work is supported by National Nature
Science Foundation of P.R. China(No. 10721063,10575126 and 10975169).
\par\vskip20pt


\end{document}